\documentclass[a4paper,11pt]{article}
\usepackage{pos}

\DeclareMathOperator*{\argmin}{arg\,min}

\newcommand{\smefit}{{\sc SMEFiT}}

\renewcommand{\vec}[1]{\textbf{#1}}

\title{The Monte Carlo replica method: investigating the effects of non-linearity}

\author*[a]{Mark N. Costantini}

\affiliation[a]{DAMTP, University of Cambridge, Wilberforce Road, Cambridge, CB3 0WA, United Kingdom\\}

\emailAdd{mnc33@cam.ac.uk}

\abstract{
This paper presents an in-depth mathematical analysis of the Monte Carlo replica method, 
commonly used in global fitting studies within the high-energy physics theory field. 
For the first time, we offer a rigorous derivation of the parameter distributions resulting from this method, demonstrating that, 
while they align with Bayesian posteriors in linear models, they deviate in non-linear cases. 
We then numerically assess this discrepancy in a phenomenologically important context: fitting SMEFT Wilson coefficients. 
Our findings reveal that when non-linearity plays a significant role, the uncertainty estimates for key quantities differ between the two approaches.    
}

\FullConference{31st International Workshop on Deep Inelastic Scattering (DIS2024)\\
 8–12 April 2024\\
Grenoble, France\\}

\begin{document}
\maketitle

\section{Introduction}
\label{sec:intro}
This paper studies an inference methodology which is widely used in the high-energy physics community, the \textit{Monte Carlo (MC) replica method}.
This method has been deployed in various fits of the Wilson coefficients in the Standard Model Effective Field Theory (SMEFT) such as ~\cite{Giani:2023gfq, Biekoetter:2018ypq}, 
and to fits of the parton distribution functions (PDFs), which parametrise hadron structure, such as ~\cite{NNPDF:2021njg, Hunt-Smith:2023sdz} 
(and, indeed, in simultaneous extractions of PDFs and SMEFT Wilson coefficients~\cite{Costantini:2024xae}).

The aim of this work is to provide a rigorous mathematical derivation of the Monte Carlo replica method, and to compare its results
with those obtained from a Bayesian approach.
We will show that the Monte Carlo replica method is typically inequivalent to the Bayesian method.
We will also provide a numerical benchmark of the Monte Carlo replica method in the context of SMEFT Wilson coefficient fits, 
and show that the Monte Carlo replica method can lead to biased and underestimated uncertainties in the parameter estimates.

\section{The mathematics of the Monte Carlo replica method}
\label{sec:multi}
In order to fix notation for the discussion of the Monte Carlo replica method in the sequel, let us suppose that experimental data comprising $N_{\text{dat}}$ datapoints is distributed according to a multivariate normal distribution:\footnote{In principle, other distributions could be considered, however, in the high-energy physics literature the multivariate normal is often used when applying the Monte Carlo replica method.}
\begin{equation}
\vec{d} \sim \mathcal{N}(\vec{t}(\vec{c}), \Sigma),
\label{eq:data_distribution}
\end{equation}
where $\Sigma$ is an $N_{\text{dat}} \times N_{\text{dat}}$ experimental covariance matrix (of which we assume perfect knowledge), and $\vec{t} : \mathbb{R}^{N_{\text{param}}} \rightarrow \mathbb{R}^{N_{\text{dat}}}$ is a smooth theory prediction function, taking as argument a vector of $N_{\text{param}}$ unknown theory parameters $\vec{c} \in \mathbb{R}^{N_{\text{param}}}$. Given an observation $\vec{d}_0$ of the experimental data, our aim is to recover a reliable estimate of the region in which $\vec{c}$ lies.

In the Monte Carlo replica method, we begin by introducing the \textit{pseudodata distribution} $\vec{d}_p \sim \mathcal{N}(\vec{d}_0, \Sigma)$, which is intended 
to approximate the actual distribution from which the measurement $\vec{d}_0$ was drawn. 
We then define the corresponding `\textit{best-fit parameter}' values to be those which minimise the $\chi^2$-statistic evaluated on 
the pseudodata:\footnote{This equation does not necessarily hold exactly in practice; often, numerical 
implementations of the Monte Carlo replica method also use a random training-validation split when finding the minimum, together with a cross-validation stopping.}
\begin{equation}
\vec{c}_p(\vec{d}_p) := \argmin_{\vec{c}} \chi^2_{\vec{d}_p}(\vec{c}) = \argmin_{\vec{c}}\ (\vec{d}_p - \vec{t}(\vec{c}))^T \Sigma^{-1} (\vec{d}_p - \vec{t}(\vec{c})),
\label{eq:monte_carlo_definition}
\end{equation}
essentially defining $\vec{c}_p(\vec{d}_p)$ as a function of a random variable $\vec{d}_p$. 
The distribution of the \textit{Monte Carlo replicas} $\vec{c}_p(\vec{d}_p)$ is now usually interpreted in the same way as 
the Bayesian posterior, in particular, by performing a change of variables we can write its distribution to be proportional to
\begin{equation}
p(\vec{c}) \propto \int d^d\vec{d}_p\ \delta( \vec{c} - \vec{c}_p(\vec{d}_p)) \exp\left( - \frac{1}{2}(\vec{d}_p - \vec{d}_0)^T \Sigma^{-1} (\vec{d}_p - \vec{d}_0) \right).
\label{eq:initial_mc}
\end{equation}
The above distribution, which is in general non-tractable, will be compared against the Bayesian posterior which is proportional to
\begin{equation}
p(\vec{c} | \vec{d}_0) \propto p(\vec{c}) \cdot p(\vec{d}_0 | \vec{c}) = p(\vec{c}) \exp\left( - \frac{1}{2} (\vec{d}_0 - \vec{t}(\vec{c}))^T \Sigma^{-1} (\vec{d}_0 - \vec{t}(\vec{c})) \right).
\label{eq:bayesian_definition}
\end{equation}

In the following section, we shall give the explicit form of the Monte Carlo posterior for a relevant toy model and compare it to the Bayesian one. 
A more detailed derivation and additional examples can be found in \cite{Costantini:2024wby}.

\subsection{Toy example: quadratic theory}
\label{subsec:toy_examples}
We present a toy model so as to illustrate more concretely the ideas discussed in the previous section. The example will also help clarify the numerical results presented in the next section.


Consider $t(c) = t_0 + t_{\text{lin}} c + t_{\text{quad}}c^2$, a quadratic theory in one parameter, with $t_{\text{quad}} > 0$, and a single datapoint measurement $d_0$ with variance $\sigma^2$. In this case, the Jacobian matrix is:
\begin{equation}
\frac{\partial t}{\partial c} = t_{\text{lin}} + 2c t_{\text{quad}},
\end{equation}
which has full rank unless $c = -t_{\text{lin}}/2t_{\text{quad}}$. 
By explicitly solving the delta function in Eq. \eqref{eq:initial_mc} and taking into account 
the degenerate solution at $-t_{\text{lin}}/2t_{\text{quad}}$, we can write the Monte Carlo posterior as
\begin{align}
2|2 c t_{\text{quad}} + t_{\text{lin}}| \exp\left( -\frac{1}{2}\frac{(d_0-t(c))^2}{\sigma^2} \right) + \delta\left( c + \frac{t_{\text{lin}}}{2t_{\text{quad}}} \right) \int\limits_{-\infty}^{t_{\rm min}} d d_p \exp\left( -\frac{1}{2}\frac{(d_0-d_p)^2}{\sigma^2} \right).
\end{align}

Importantly, this demonstrates a singular behaviour of the Monte Carlo posterior at the point $c = -t_{\text{lin}}/2t_{\text{quad}}$. Indeed, this can result in a significant bias in the central value and a significant underestimation of the uncertainties for the parameter $c$, since the posterior is highly concentrated around $c = -t_{\text{lin}}/2t_{\text{quad}}$. This phenomenon is showcased in the phenomenologically relevant case of the SMEFT in the next section.

\section{Applications to high energy physics}
\label{sec:applications}
In this section, we apply the mathematical discussion from Section~\ref{sec:multi} to a phenomenological example in high-energy physics. In particular, we contrast the Bayesian and Monte Carlo posteriors in the case of fits of SMEFT Wilson coefficients. Note that a more detailed discussion and the same exercise in the context of uncertainty estimation for the PDFs of the proton is discussed in Sect. (3) of \cite{Costantini:2024wby}.

\paragraph{SMEFT fits} The \textit{Standard Model Effective Field Theory} (SMEFT) treats the SM as a low-energy effective limit of an ultraviolet theory. As such, it extends the SM Lagrangian by a series of non-renormalisable operators built from the SM fields and respecting the SM symmetries:
\begin{equation}
\mathcal{L}_{\text{SMEFT}} = \mathcal{L}_{\text{SM}} + \sum_{d=5}^{\infty} \sum_{i=1}^{N_d} \frac{c_d^{(i)} \mathcal{O}_d^{(i)}}{\Lambda^{d - 4}},
\end{equation}
where $\Lambda$ is some characteristic energy scale of New Physics 
, and the sum over $d$ is a sum over the dimension of the operators. At each dimension $d$, there are $N_d$ independent operators, $\mathcal{O}_d^{(i)}$, indexed by $i=1,...,N_d$, with corresponding couplings $c_d^{(i)}$. The couplings $c_d^{(i)}$ are called \textit{Wilson coefficients}, and parametrise deviations from the SM.

Cross-section predictions from the dimension-six SMEFT generically take the form:
\begin{equation}
\vec{t}(\vec{c}) = \vec{t}_{\text{SM}} + \vec{t}_{\text{lin}} \vec{c} + \vec{t}_{\text{quad}} (\vec{c} \otimes \vec{c}),
\end{equation} 
where $\vec{t}_{\text{SM}}$ is a $N_{\text{dat}} \times 1$ vector of SM predictions, $\vec{t}_{\text{lin}}$ is an $N_{\text{dat}} \times N_{\text{op}}$ matrix of linear SMEFT predictions (with $N_{\text{op}}$ the number of SMEFT operators in the fit), and $\vec{t}_{\text{quad}}$ is an $N_{\text{dat}} \times N_{\text{op}}^2$ matrix of quadratic SMEFT predictions. 
Importantly, this is a non-linear theory, and as we saw in the example of Sect.~\ref{subsec:toy_examples} this can lead to discrepancies between the Monte Carlo posterior and the Bayesian posterior (unless a specific highly non-trivial prior is chosen). The quadratic behaviour of the SMEFT theory predictions additionally implies that there may be delta function singularities as we saw in Example 2 of Sect.~\ref{subsec:toy_examples}.  

We perform an analysis of the 175 measurements used in the simultaneous PDF and SMEFT fit of Ref.~\cite{Kassabov:2023hbm}, encompassing $t \bar{t}$, $t \bar{t} + X$, single top, single top $+X$, $t \bar{t} t \bar{t}$, $t \bar{t} b \bar{b}$ and top decay observables.  These observables will be used to determine the posterior distributions of 25 Wilson coefficients of the dimension-6 SMEFT, the notation for which can be found in Table B.1 of~\cite{Kassabov:2023hbm}. The \smefit{} code~\cite{Giani:2023gfq} is used to compare the Nested Sampling and Monte Carlo replica methodologies due to the availability of both methodologies in the public code.

\begin{figure}[htb!]
\centering
\includegraphics[scale=0.25]{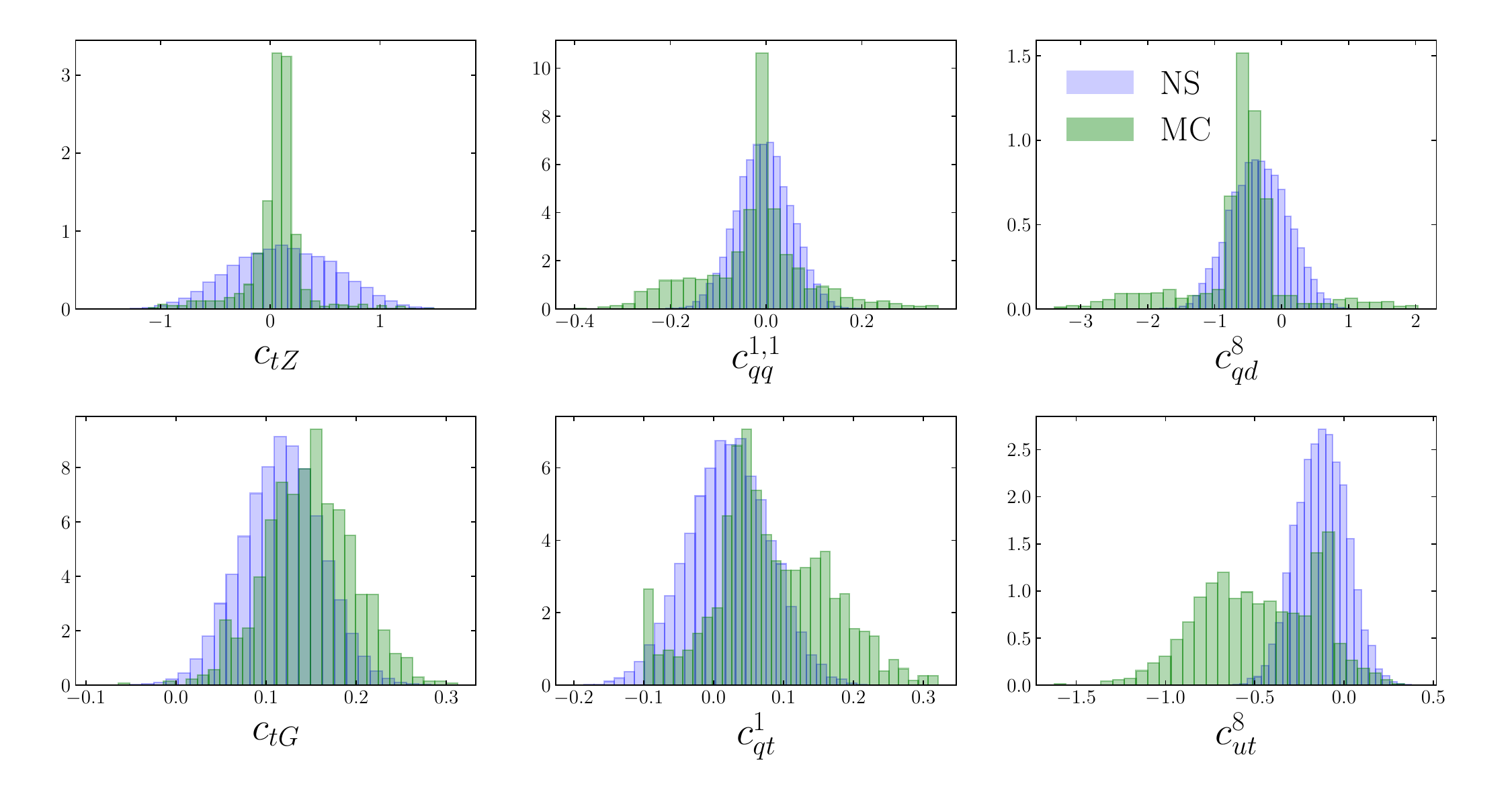}
\caption{Selection of SMEFT posteriors showcasing significant differences in the results obtained by Nested Sampling and the Monte Carlo replica method.}
\label{fig:smeft_posteriors_comparison}
\end{figure}

In Fig.~\ref{fig:smeft_posteriors_comparison} we compare the Bayesian and Monte Carlo posteriors for some of the fitted Wilson coefficients.
For $c_{tZ}$, in particular, we observe a narrow constraint for the Monte Carlo replica method due to a spiked distribution,
which showcases the delta function behaviour discussed in Sect.~\ref{subsec:toy_examples}. Similarly spiked behaviour is observed in the posterior distributions of the four-fermion operators, in particular $c_{qq}^{1,1}$ and $c_{qd}^{8}$, while the posterior distributions of $c_{qt}^{1}$ and $c_{ut}^{8}$ are found to be highly skewed and non-Gaussian relative to the Bayesian posteriors.
Finally, we observe that for $c_{tG}$, both methodologies obtain Gaussian-like posterior distributions of similar widths; however the Monte Carlo replica method shifts the distribution in the positive direction, leading to a stronger pull from the SM than that obtained with Nested Sampling.

\section{Conclusions}{}
\label{sec:conclusions}
For the first time, this work explored the mathematical foundations of the Monte Carlo replica method beyond linear models.

In Sect.~\ref{sec:multi}, we outline the main steps for the calculation of the Monte Carlo `posterior' distribution and show that in general it does not 
agree with the Bayesian method. Some of the behaviour of the Monte Carlo replica posterior has been understood (particularly the delta function singularities that can appear in the posterior), but for the most part it remains intractably difficult to manipulate and requires a numerical analysis.

In Sect.~\ref{sec:applications}, we provided a benchmark of the Bayesian method against the Monte Carlo replica method in the context of a fit of the SMEFT Wilson coefficients in the top sector and found poor agreement between the implied parameter bounds.

We believe that further efforts should be dedicated towards the development of a fully Bayesian approach in non-linear inverse problem in high energy
physics, a notable example of which could be that of PDF fitting, especially if fit simultaneously in combination with EFT coefficients.
This work offers a compelling motivation for such a programme.

\renewcommand{\em}{}
\bibliographystyle{UTPstyle}
\bibliography{references}

\providecommand{\href}[2]{#2}\begingroup\raggedright\begin{thebibliography}{1}

\bibitem{Giani:2023gfq}
T.~Giani, G.~Magni, and J.~Rojo, {\it {SMEFiT: a flexible toolbox for global
  interpretations of particle physics data with effective field theories}},
  \href{http://arxiv.org/abs/2302.06660}{{\tt arXiv:2302.06660}}.

\bibitem{Biekoetter:2018ypq}
A.~Biekoetter, T.~Corbett, and T.~Plehn, {\it {The Gauge-Higgs Legacy of the
  LHC Run II}},  {\em SciPost Phys.} {\bf 6} (2019), no.~6 064,
  [\href{http://arxiv.org/abs/1812.07587}{{\tt arXiv:1812.07587}}].

\bibitem{NNPDF:2021njg}
{\bf NNPDF} Collaboration, R.~D. Ball et~al., {\it {The path to proton
  structure at 1\% accuracy}},  {\em Eur. Phys. J. C} {\bf 82} (2022), no.~5
  428, [\href{http://arxiv.org/abs/2109.02653}{{\tt arXiv:2109.02653}}].

\bibitem{Hunt-Smith:2023sdz}
{\bf Jefferson Lab Angular Momentum (JAM)} Collaboration, N.~T. Hunt-Smith,
  W.~Melnitchouk, N.~Sato, A.~W. Thomas, X.~G. Wang, and M.~J. White, {\it
  {Global QCD analysis and dark photons}},  {\em JHEP} {\bf 09} (2023) 096,
  [\href{http://arxiv.org/abs/2302.11126}{{\tt arXiv:2302.11126}}].

\bibitem{Costantini:2024xae}
M.~N. Costantini, E.~Hammou, Z.~Kassabov, M.~Madigan, L.~Mantani,
  M.~Morales~Alvarado, J.~M. Moore, and M.~Ubiali, {\it {SIMUnet: an
  open-source tool for simultaneous global fits of EFT Wilson coefficients and
  PDFs}},  \href{http://arxiv.org/abs/2402.03308}{{\tt arXiv:2402.03308}}.

\bibitem{Costantini:2024wby}
M.~N. Costantini, M.~Madigan, L.~Mantani, and J.~M. Moore, {\it {A critical
  study of the Monte Carlo replica method}},
  \href{http://arxiv.org/abs/2404.10056}{{\tt arXiv:2404.10056}}.

\bibitem{Kassabov:2023hbm}
Z.~Kassabov, M.~Madigan, L.~Mantani, J.~Moore, M.~Morales~Alvarado, J.~Rojo,
  and M.~Ubiali, {\it {The top quark legacy of the LHC Run II for PDF and SMEFT
  analyses}},  {\em JHEP} {\bf 05} (2023) 205,
  [\href{http://arxiv.org/abs/2303.06159}{{\tt arXiv:2303.06159}}].

\end{thebibliography}\endgroup

\end{document}